\documentclass[runningheads]{llncs}
\usepackage{graphicx}
\usepackage{booktabs}
\begin{document}
\title{An Investigation into Keystroke Dynamics and Heart Rate Variability as Indicators of  Stress }
\titlerunning{Keystroke dynamics as stress indicators}
%
\author{Srijith Unni, Sushma Suryanarayana Gowda,
Alan F. Smeaton\orcidID{0000-0003-1028-8389}}
\authorrunning{S. Unni et al.}
%
\institute{Insight Centre for Data Analytics\\
Dublin City University, Glasnevin, Dublin 9, Ireland\\
\email{alan.smeaton@dcu.ie}\\
}
\maketitle              
\begin{abstract}

Lifelogging has become a prominent  research topic in recent years. Wearable sensors like Fitbits and smart watches are now  increasingly popular for recording one's activities. Some researchers are also exploring 
key\-stroke dynamics for lifelogging. Keystroke dynamics refers to the process of measuring and assessing a person’s typing rhythm on digital devices. A digital footprint is created when a user interacts with devices like keyboards, mobile phones or touch screen panels and the timing of the keystrokes is  unique to each individual though likely to be affected  by  factors such as  fatigue,  distraction or emotional stress.  In this work we  explore the relationship between  keystroke dynamics as measured by the timing for the top-10 most frequently occurring bigrams in English, and the emotional state and stress of an individual as measured by heart rate variabiity (HRV). We collected  keystroke data using the Loggerman  application  while HRV was simultaneously gathered. With this data we performed an analysis to determine the relationship between  variations in keystroke dynamics and variations in HRV. Our conclusion is that we need to use a more detailed representation of keystroke timing than the top-10 bigrams, probably personalised to each user.

\end{abstract}

\section{Introduction}

There are multiple ways in which we can capture aspects of our lives to record our  activities and even our state of mind. Typically this requires wearable devices for measuring aspects of our physiology (heart rate, respiration rate), our activities (step counters and location trackers) or our environment (wearable cameras, bluetooth detectors).  The dependency is on using  wearable devices but the body is a hostile environment for wearables and we are not always comfortable wearing them, such as when we sleep for example.  Thus environmental sensing can  substitute for wearables and  {\em in situ} sensing in a home environment can tell as much about  occupants' activities as wearables can.

In this work  we  use another form of environmental sensing -- keystroke dynamics -- to explore what we can infer about the mental state and mental stress on a user at a point in time. 
Keystroke dynamics or typing dynamics refers to the detailed timing information which describes exactly when each key was pressed and when it was released, as a person is typing on a keyboard. Keystroke dynamics are unique to each individual and is known to vary according to factors including fatigue, stress, and emotional state \cite{crawford2010keystroke}. 
We capture information on user state using keystroke dynamics as a form of contextual data. As a ground truth against which to measure, we use heart rate variability (HRV), a measure of timing variations between two consecutive heartbeats  known as an indicator of mental strain, stress and mental workload \cite{hjortskov2004effect}.  If we can correlate data drawn from keystroke dynamics with variations in HRV then we will  demonstrate that using keystroke dynamics we can measure some aspects of a user's mental state or stress.

Research on keystroke dynamics applications has been increasing, especially in the last decade. The main motivation behind this is the uniqueness of the patterns to an individual person and the ease of data collection. Keystroke events can now be measured up to milliseconds precision by software \cite{senk2011biometric}.  In contrast to traditional physiological biometric systems such as palm print, iris and fingerprint recognition that rely on dedicated device and hardware infrastructure, keystroke dynamics recognition is entirely software implementable. The benefit of low dependency on specialised hardware not only can significantly reduce deployment cost but also creates an ideal scenario for implementation in remote authentication environments \cite{teh2013survey}.

The approach we take here is to collect  keystroke data from  subjects  using a simple keystroke logging application called Loggerman. Loggerman \cite{hinbarji2016loggerman} is an application which collects data generated from normal usage of a computer system  including keystrokes data and it operates passively in the background. For the collection of physiological data  we   use    heart rate variability, which we use as a proxy for emotional state and which we introduce in the next section.

In the next section we review related literature on uses of keystroke dynamics and the section following that presents an overview of heart rate variability. We then describe the data we have gathered from participants and how this was prepared and analysed, and the results we have obtained.

\section{Literature Review}

The earliest noted use of keystroke dynamics was to identify individual telegraph operators in late 19th century by listening to the patterns of their taps on the device \cite{teh2013survey}. Analysis of keystroke dynamics was explored in the 1970s with the study of time intervals between keystrokes and other keying rate information \cite{neal1977time}, with further research focusing on how this information can be applied for practical purposes. One of the main aspects in which prospective applications were researched  was user authentication and interpreting how different users can be identified from their typing patterns  \cite{bergadano2002user}.

The behavioural biometrics of Keystroke Dynamics uses the manner and rhythm in which an individual types characters on a keyboard or keypad \cite{deng2013keystroke}. The keystroke rhythms of a user can be measured to develop a unique biometric template of a user's typing patterns for future authentication \cite{panasiuk2010modified}. Keystrokes are separated into static and dynamic typing, which are used to help distinguish between authorised and unauthorised users. Additionally, integration of keystroke dynamics biometrics leaves random password guessing attacks obsolete \cite{de1997enhanced}.

In \cite{leijten2013keystroke},  Leijten \emph{et al.}  described keystroke logging as an instrumental method in determining writing strategies and established how cognitive actions are performed and correlate with keystroke timings. 
One  reason for this that has been identified is that keystroke timing data reveals various ebbs and flows in the fluency of writing which can be interpreted and can help further understand the cognitive process going on as we write. Gunetti \emph{et al.} \cite{gunetti2005keystroke} concluded in their research that ``typing dynamics is the most natural kind of biometrics that stems from the use of computers, it is relatively easy to sample, and it is available throughout the entire working session.”

The ability to recognise emotions is an important part of building intelligent computers. Emotionally-aware systems  have a rich context from which to make appropriate decisions about how to interact with the user or adapt their system response \cite{epp2011identifying}. Various algorithms like facial emotional analysis, auditory emotional analysis, sentiment analysis, and emotional body gesture recognition are currently used to determine human emotions. The problem with these  approaches for identifying emotions that limit their applicability is that they are invasive and  can require costly equipment \cite{epp2011identifying}. 

Teh \emph{et al.} \cite{teh2013survey} discussed in their paper that although there certain advantages such as uniqueness, transparency and non-invasiveness, keystroke dynamics can reveal variations in typing rhythm  caused by external factors such as injury, fatigue or distractions. Hence there is an opportunity to investigate keystroke dynamics to indicate the mental state of a person at a point in time. For example, a person might compose an email quicker when s/he realises they would get off work right after that action, compared to a person who knows s/he must work hours after that mail has been composed. Such differences in the timing of  typing, if logged, would indirectly enable us to identify certain models to determine the emotional state or stress levels in a more generalised manner. We believe that this could help us observe how emotion correlates with the keystroke data. Vizer \emph{et al.} \cite{vizer2009automated} confirmed in their research ``the potential of monitoring keyboard interactions for an application other than security and highlighted the opportunity to use keyboard interaction data to detect the presence of cognitive or physical stress.”

Some previous research has been reported to determine the emotions of a user using keystroke dynamics.  In \cite{epp2011identifying}, Epp \emph{et al.} determine user emotions by comparing the rhythm of  typing patterns on a standard keyboard with emotional state as collected via self-report. They claim that the results include 2-level classifiers for confidence, hesitance, nervousness, relaxation, sadness, and tiredness with accuracy ranging from 77 to 88\%. In addition, they show promise for anger and excitement, with accuracy of 84\%. One problem with this approach is that the emotional state of their subjects is collected using self-reporting. This is not completely reliable since humans have a tendency to mis-report, for various reasons. 
More recently, \cite{smeaton2021keystroke},  analysed variations in keystroke dynamics with respect to the previous day’s sleep score for participants but could not find any significant relationship between the two.

\section{Heart Rate Variability (HRV)}

Heart rate variability (HRV) is a measure of  variations observed between two consecutive heartbeats. HRV originates in the autonomic nervous system that is responsible for  involuntary aspects of our physiology. The autonomic nervous system has two branches, the parasympathetic system (deactivating/rest) which handles inputs from internal organs and causes decrease in heart rate, and the sympathetic system (activating/fight or flight) which handle inputs from external factors like fatigue, stress, exercise and  increases the heart rate. If our nervous system is balanced, the heart is constantly told to beat faster by the sympathetic system and  slower by the parasympathetic system. Thus there is a fluctuation caused in the heart rate and this is   HRV \cite{hjortskov2004effect}.

A high measure of heart rate variability means that our body is responsive to both sets of inputs (parasympathetic and sympathetic). This is a sign that our nervous system is balanced, and that our body is very capable of adapting to its environment and performing at its best. On the other hand, if we have a low heart rate variability, one branch is dominating (usually the sympathetic) and sending stronger signals to our heart than the other branch. However, if we are not doing something active and we have a low heart rate, then a low HRV at such periods indicates that our body is working hard for some other reason, perhaps fatigue, dehydration, stress, or we are ill and need to recover, and that leaves fewer resources available to dedicate towards other activities like exercising, competing, giving a presentation at work, etc.

\begin{figure}[!htb]
\begin{center}
\includegraphics[width=0.6\textwidth]{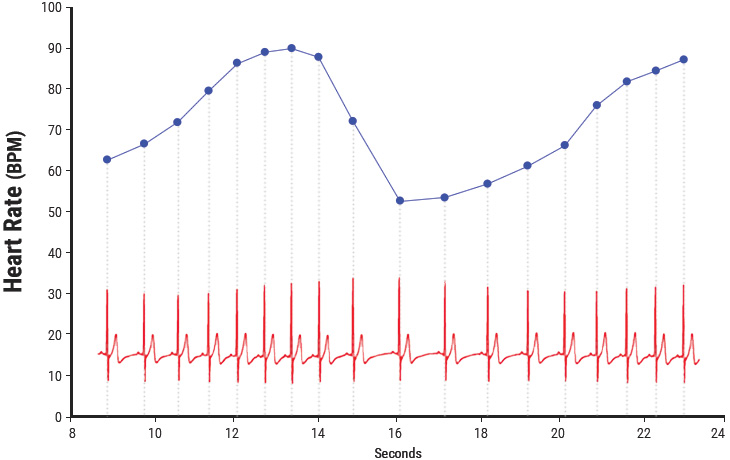}
\end{center}
\caption{Heart Rate Variability derived from Heart Rate measurements
\label{fig:heart-rate-variabilty}}
\end{figure}

\noindent 
In Figure~\ref{fig:heart-rate-variabilty}, we  observe an electrocardiogram (ECG) which is shown as the bottom graph, and the instantaneous heart rate is shown at the top. If we examine the time taken between each of the heartbeats (the blue line) between 0 and approximately 13 seconds, we notice that they become progressively shorter as heart rate accelerates and start to decelerate at around 13 seconds. Such a pattern of heart-rate acceleration and deceleration is the basis of the heart's rhythms \cite{noauthor_chapter_nodate}.

Acharya \emph{et al.} in \cite{acharya2006heart} explained that there is a clear relationship between HRV and gender and age and that physically active young and old women have higher HRV whereas lower HRV is observed in  alert new-born males compared to females. In a study among healthy subjects between 20 and 70 years of age, it was observed that HRV decreases with age and that variations in HRV are less in the case of men. 

The method for calculating HRV is as follows. A time domain index uses statistical measures to quantify the amount of variance in the inter-beat interval (IBI) of a heart beat.  The three most important and frequently reported time-domain metrics are SDNN, SDNN index, and RMSSD \cite{noauthor_chapter_nodate}. SDNN is the standard deviation of the NN intervals of  normal sinus beats as measured in milliseconds where  normal involves removing abnormal beats like ectopic beats. The NN interval, sometimes known as the R-R interval, is the time difference between successive heartbeats measured from the point of peak voltage amplitude. The standard short-term recording of HRV is of 5 minutes duration.  

HRV is known as an indicator of mental strain, stress,  and mental workload~\cite{kaur2014heart} and thus as a quantitative metric is has strong value.
The traditional method of collecting data from short-term (5-minute) heart rate  series involves a dedicated electrocardiograph (ECG) running from a computer or micro-controller and connected to a single participant in a laboratory environment. The HRV value is calculated based on the R wave time series provided by the ECG, which is the signal history of ventricular depolarisation over time. This collection procedure is  common in published studies but this experimental setup does not suit our needs.

Reliable HRV calculation outside the laboratory faces many   challenges, though recently the quality of wearable sensing has made this feasible. Optical sensors for measuring heart rate are inexpensive, portable, computationally efficient,  non-invasive, reusable, and have low power consumption so as wearable devices they  have a long time between charges. Thus optically derived pulse rate is an ideal choice for 
wearables to collect a good estimation  of  HRV value.

For our work we  use a Scosche RHYTHM24 fitness band to capture heartbeat data for our test subjects. Data recorded by the band is passed to the SelfLoops HRV mobile app  for further analysis. In the next section we describe our data collection.

\section{Data Gathered}

Ethical approval for this work was granted by the Ethics Committee of the School of Computing at Dublin City University and consisted of  a \emph{Plain Language statement}  and an \emph{Informed Consent Form} signed by  participants indicating their understanding and approval. Participants were  provided with an instruction manual for collecting both their keystroke data   with the Loggerman application on their Mac and  HRV data  collected via the Scosche RHYTHM24 fitness band.  Participants were asked   to wear the band at the appropriate position on their arm  and to start recording their HRV data at least 30 minutes before they start using  their laptop.  

Participants used Loggerman   and HRV data collection while using their laptops for planned typing ``episodes'' of long duration. For example, just quickly checking email or news first thing in the morning would not be recorded as it would not  require much typing. A planned work session to work on a paper, or write a blog, or process and respond to a stack of emails are examples of planned typing sessions which are of long enough duration and  involve typing. 
Watching a movie on a laptop would not be recorded as there is no user typing involved. 

While recording, users proceeded with their work as usual for as long as they wanted to on their Macs but the typing was not continuous as breaks would be taken for rest or refreshment, interruptions would happen because people are working from home, and thinking time for reflection or interaction with others would be interspersed with actual typing. Thus while  HRV recording is continuous, the recording of keystroke dynamics during the recording sessions is scattered and bursty, which is normal typing behaviour for most people.

After every recorded session,  participants shared their keystrokes and HRV files and the data they gathered is shown in Table~\ref{tab-users}. This shows the total number of keystrokes pressed, the number of typing episodes or instances  and the number of hours of heart rate and typing data that was gathered. The duration of the recorded sessions varied from 33 minutes to almost 7 hours for participant 1 and 25 minutes to 2 hours 45 minutes for participant 2.

\begin{table}[ht]
    \centering
    \begin{tabular}{lllll}
    \toprule
          Participant~~~~ & Keystrokes~~~~ & Typing `episodes'~~~~ & Hours of HR ~~~ & Hours of typing  \\
          \midrule
         1 &  33,583 & 17 & 48 & 23\\
         2 &    16,286 & 5 & 8 & 2.5\\
         \bottomrule
    \end{tabular}
    \caption{Details of keystroke and heart rate data logged by participants}
    \label{tab-users}
\end{table}

\section{Data Preparation}

Keystroke dynamics data  gathered by  Loggerman is in the form of a  code for each key pressed,
and a   Unix timestamp measured in milliseconds. A major challenge was to extract features from this raw time series  data so we can identify variations in keystroke dynamics across time,  for each participant. 


Peter Norvig published results of an analysis of the letters, words and n-gram frequencies  extracted from the Google Books collection of 743,842,922,321 word occurrences in the English Language. 
In this, Norvig observed the top 10 most frequently occurring bigrams in the English language to be the following - TH, HE, IN, ER, AN, RE, ON, AT, EN and ND.\footnote{http://norvig.com/mayzner.html} 
Given that these 10 bigrams are likely to occur frequently in the text typed by our participants we  use the time taken to type these 10 bigrams, the number of milliseconds between typing the two characters, as  features for characterising a typing episode.


Our data pre-processing was performed using Python libraries on Google Colab. As our analysis focused  on the top 10 bigrams observed by Norvig's analysis,  we identified    instances of each bigram  for  recorded sessions but only where the characters were typed within a 1000ms window. Our reasoning is that if it takes longer than 1 second between typing these bigram characters then there is an interruption to the flow of typing and we are interested in episodes of continuous typing.

\begin{figure}[!htb]
\begin{center}
\includegraphics[width=\textwidth]{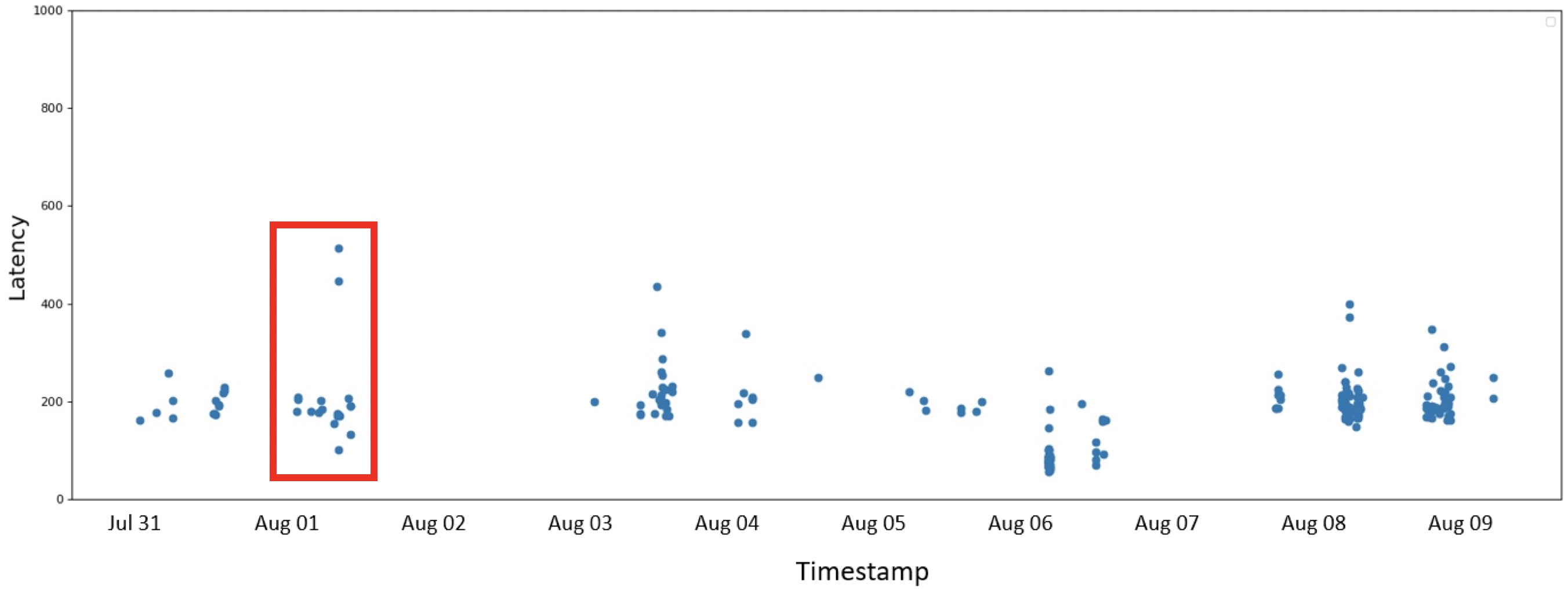}
\end{center}
\caption{Sample of typing episodes observed for typing of the ER bigram}
\label{fig:ER and ND Bigram}
\end{figure}

Figure~\ref{fig:ER and ND Bigram} shows a snapshot of some of the typing episodes for the ER bigram chosen as an illustration, which were  observed over 10 days for one participant. The graph  shows observed typing episodes of differing durations with gaps removed, so it does not display 24 hours per day. We  observe   distinct episodes  where the  bigram was used considerably rather than  rarely. The red box on Aug 01 highlights 17 occurrences (blue dots in the figure) of an ``ER'' during one typing session with the timing for this, i.e. the latency between the keystrokes, varying from about 100ms to 550ms. We  represent this typing episode as the mean of the 17 instances, a value of around 220ms, and  repeat this computation for the other 9 of the top-10 bigrams but as we see when we look at HRV values, there can be a lot of variety of HRV and thus of stress and mental workload within a typing session thus we need more temporally fine-grained analysis of keystrokes within each typing episode.

\begin{figure}[!htb]
\begin{center}
\includegraphics[width=0.5\textwidth]{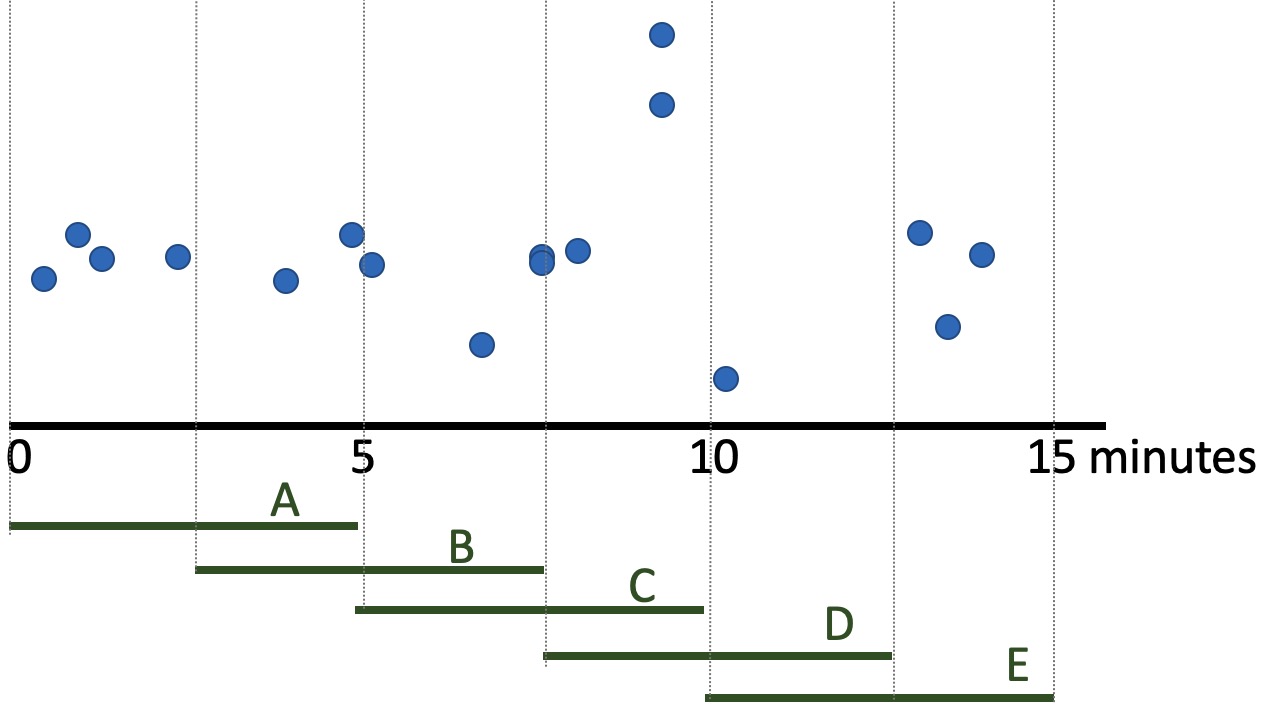}
\end{center}
\caption{Calculating keystroke timing for ER bigram in 5-minute overlapping windows}
\label{fig:schematic}
\end{figure}

Figure~\ref{fig:schematic} presents a schematic of how we do this. Using the highlighted typing episode from Figure~\ref{fig:ER and ND Bigram} which we see is of 15 minutes duration, we divide this into 5 overlapping 5-minute windows shown as green bars labelled A, B, C, D, E and for each  we compute the average latency for  ER within that window. If any 5-minute windows have no ER bigrams, they are removed from the analysis. In the case of this example we have 5 mean latencies for the typing of ER and we repeat this for all top-10 bigrams.

As outlined earlier, to calculate HRV we used short term HRV calculated from the RR-intervals, for a duration of 5 minutes. Outlier HRV values, also called   ectopic beats, were removed using Malik's rule \cite{acar2000automatic}  with the help of the HRV-analysis library in Python.
A number of 5-minute HRV values were   calculated for each of the typing sessions using the same 5-minute sliding windows with 2.5 minutes overlap for correlation with the keystroke dynamics for the same 5-minute windows.

\begin{figure*}[!htb]
\begin{center}
\includegraphics[width=\textwidth]{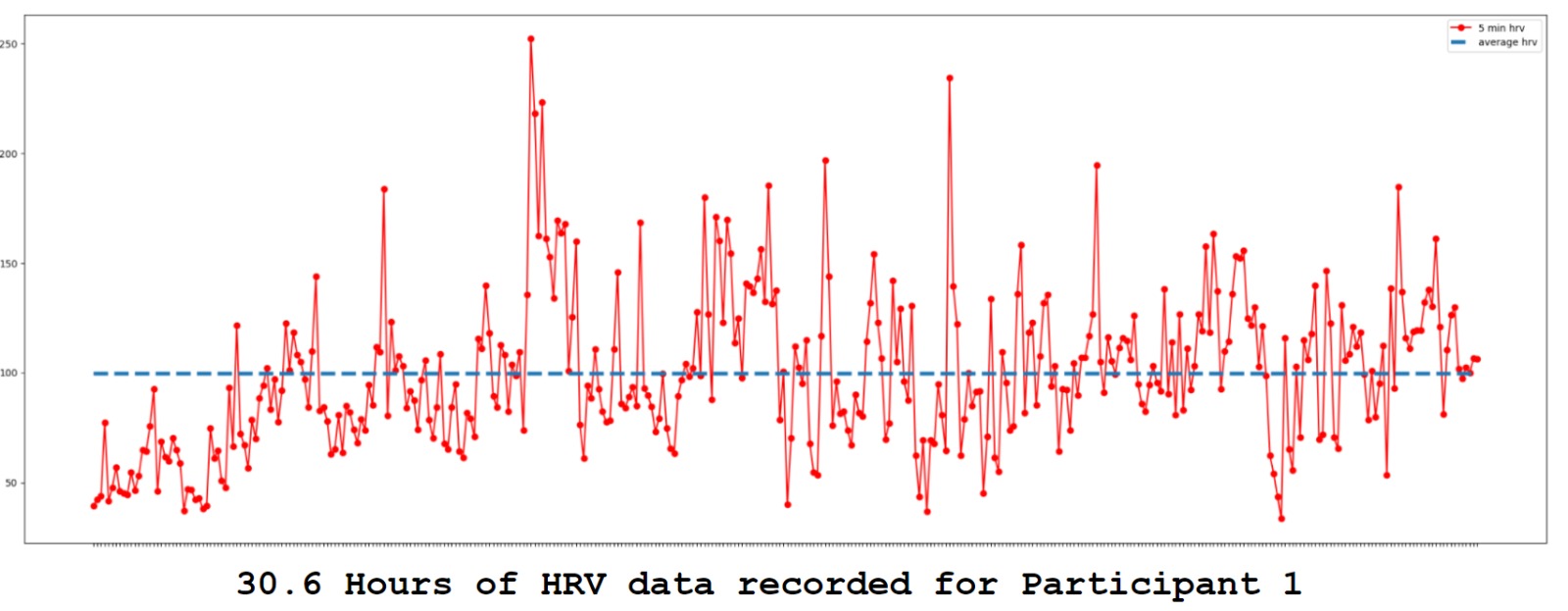}
\includegraphics[width=\textwidth]{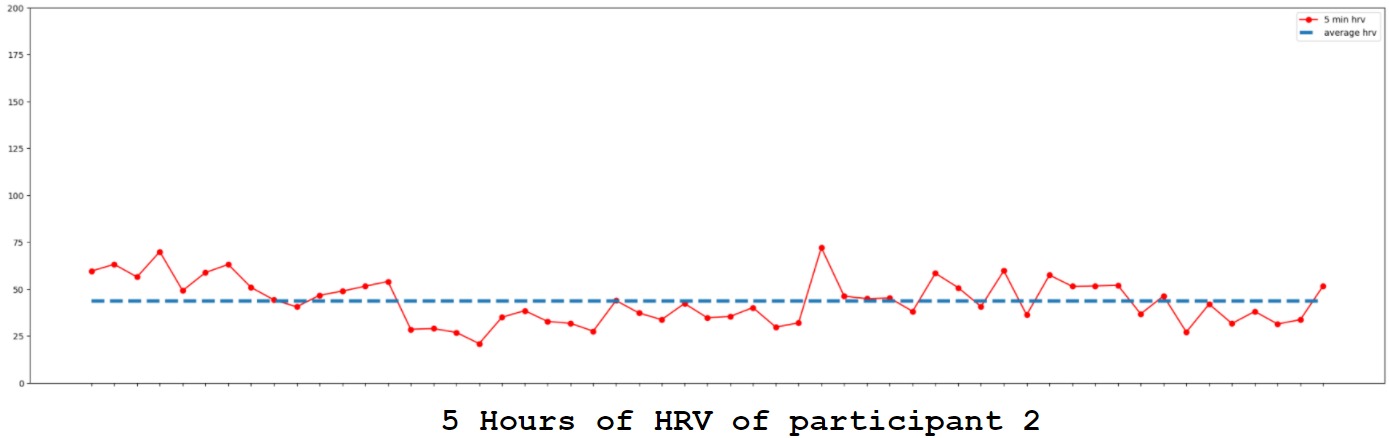}
\end{center}
\caption{A sample of the 5-minute  cumulative and contiguous HRV values for 2 participants showing how HRV varies from person-to-person and within an individual person. The top graph shows 30.6 hours and the lower shows only 5 hours.}
\label{fig:HRV values}
\end{figure*}

Using  data for all recorded sessions, for each participant we  calculated their average HRV that acts as a their HRV baseline. Figure~\ref{fig:HRV values} shows some of those (non-overlapping) HRV values for 2  participants with the dotted line showing their average HRV value. This
indicates that 
participant
1 has a lot of variability from a low of about 20ms to a high of almost 250ms whereas the variations are almost flat for Participant 2 whose range is 25ms to about 75ms. In turn this suggests that participant 1 was, at the times of logging, experiencing a range of stress levels which made their HRV values vary considerably.

\section{Data Analysis and Results}

We processed keystroke timing data by using only  the non-empty 5-minute windows within each of the  typing episodes where at least one of the top-10 bigrams had been typed within 1,000ms.   For each window we calculated the mean latency for each, for each of the ten bigrams.   We then computed the difference between those values for each bigram and the mean bigram latencies for the whole recording  for each participant which we define as their baseline, to compute the deviation  from the baseline, for each window.
We then used the simple average of all the deviation ratios  of all bigrams for the same 5-minute window as the observed change in keystroke dynamics, which we can measure against observed changes in HRV.

\begin{figure*}[!htb]
\begin{center}
\includegraphics[width=\textwidth]{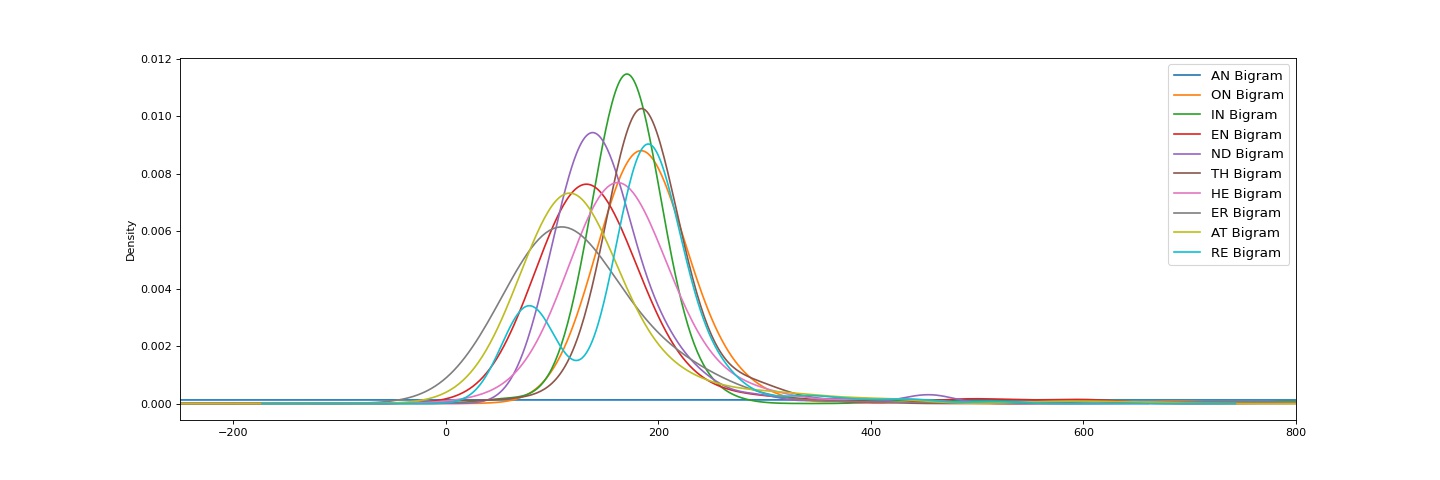}
\end{center}
\caption{Cumulative Density Plots for top 10 bigrams for collected data}
\label{cumulative density}
\end{figure*}

We can see the cumulative variations in  latencies for all bigrams  in Figure~\ref{cumulative density} and this illustrates  how  participants' typing speeds for the same top-10 bigrams vary within and across multiple typing sessions.  Table~\ref{tab:timings} shows the baseline timings for the top-10 bigrams for both users.  In practice only 44\% of the 645 5-minute windows for participant 1 had any typing of any of the top-10 bigrams and on average those windows had 4.4 of those top-10. For user 2, 71\% of the 5-minute windows had at least 1 of the top-10 bigram, on average  7.1. 

\begin{table}[ht]
    \centering
    \begin{tabular}{l|llllllllll|l}
    \toprule 
Bigram~~~ &~~TH~~~~~~&	HE~~~~~~&	IN~~~~~~&	ER~~~~~~&	AN~~~~~~&	RE~~~~~~&	ON~~~~~~&	AT~~~~~~&	EN~~~~~~&	ND~~~~~~& ~~Average \\
\midrule 
User 1 & ~~~~88.9&	~~69.5&	~~89.0&	~~56.1&	~~78.6&	475.6&	~~58.4&	~~65.4&	~~83.4&	103.9&    117 ms\\
User 2 & ~~124.0&	118.7&	141.3&	130.3&	122.9&	~~51.8&	~~82.0&	112.1&	~~76.0&	131.9&    109 ms\\
\bottomrule  
    \end{tabular}
    \caption{Baseline top-10 bigram timings (in ms)}
    \label{tab:timings}
\end{table}

\noindent 
We then calculated the HRV values for the same 5-minute overlapping windows for each of the typing sessions, for each participant.
For User 1 there were 275x overlapping 5-minute windows with HRV recorded and those HRV values varied from 46ms to 240ms with an overall average of 106.3ms.
For user 2 there were 86x overlapping 5-minute windows with HRV recorded and those HRV values  varied from 21ms to 67ms with an overall average of 40.9ms, considerably lower than for user 1.  

\begin{figure*}[!htb]
\begin{center}
\includegraphics[width=1.0\textwidth]{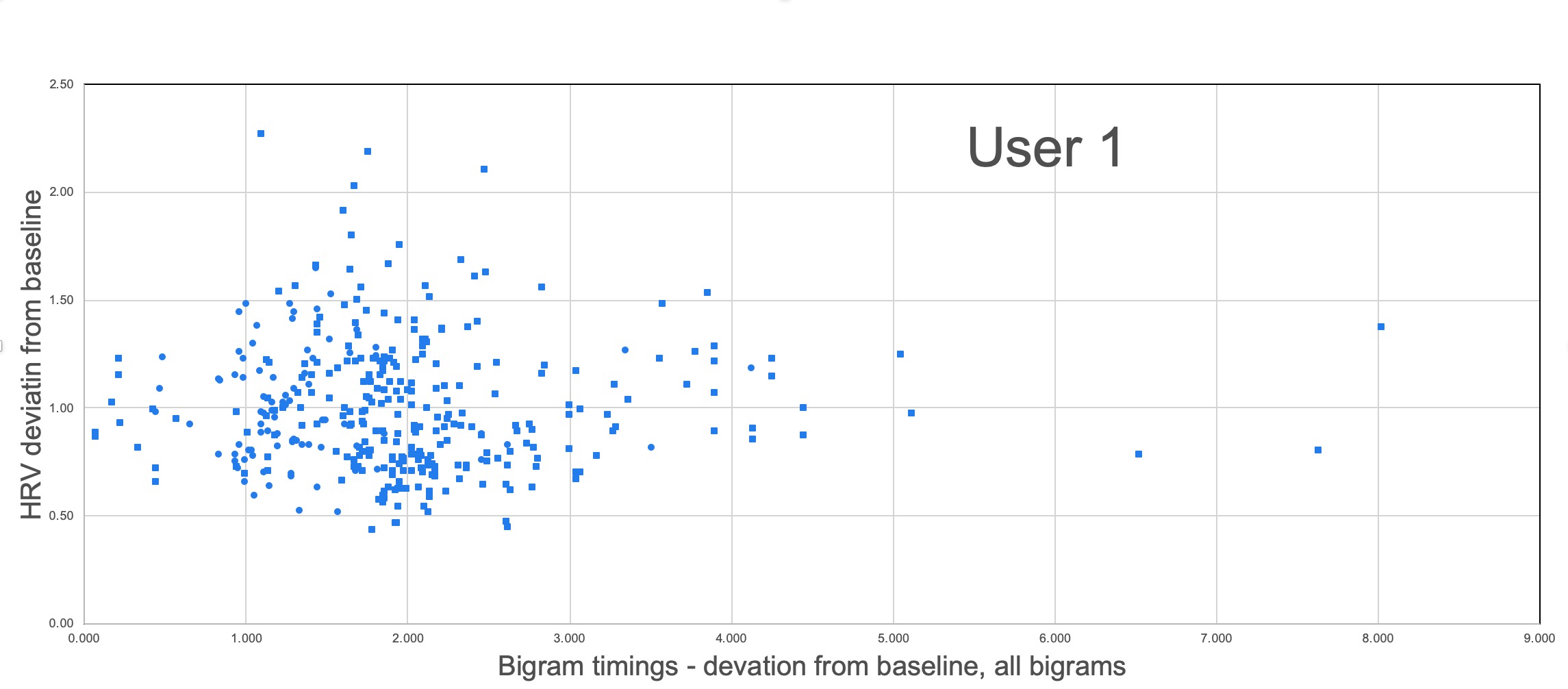}
\includegraphics[width=1.0\textwidth]{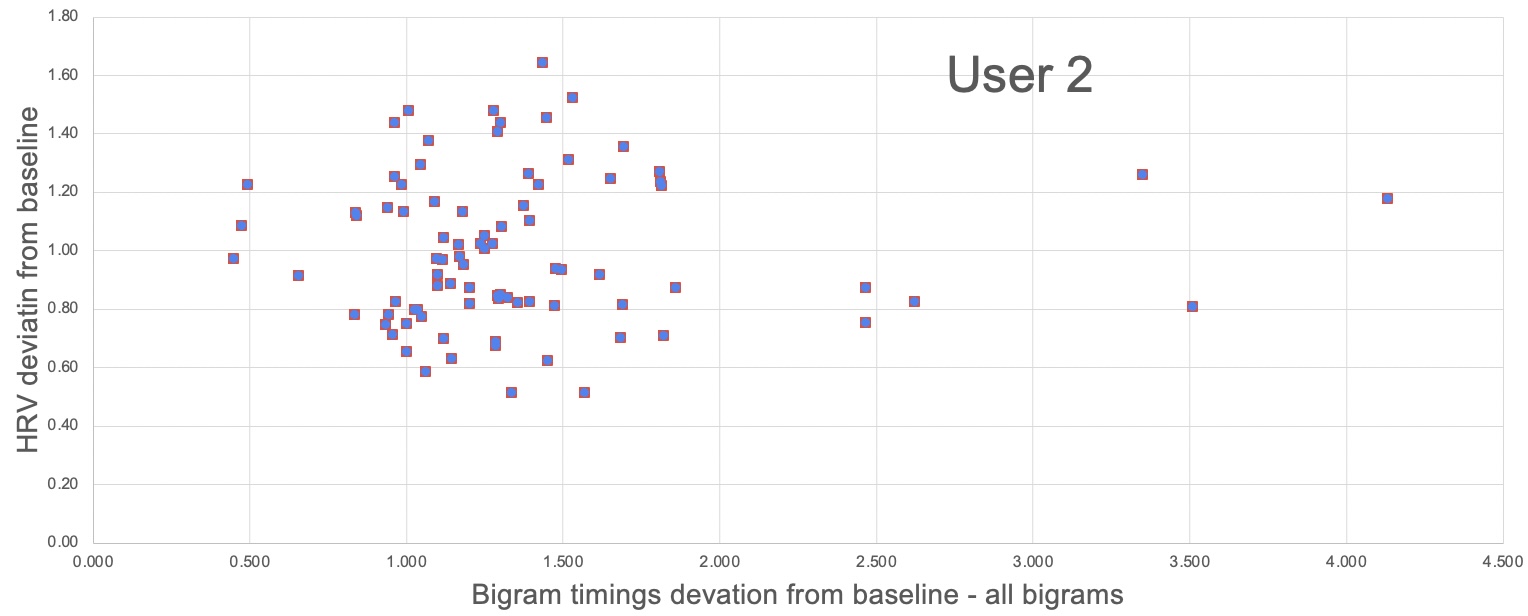}
\end{center}
\caption{Correlation between latency for timing of top-10 bigrams vs. mean  HRV for each 5-minute window, for user 1 and user 2}
\label{fig:correlation}
\end{figure*}

With  baseline HRV values and observed HRV values for each 5-minute window we can compare these against the variance between observed and  baseline keystroke timing data for each user and a scatterplot of this for user 1 is shown in  Figure~\ref{fig:correlation}.  

Figure~\ref{fig:correlation} shows there is little correlation between deviations in HRV and deviations in keystroke dynamics when all of the top-10 occurring bigrams are considered and combined, for either user.  We know user 1 has a higher baseline HRV compared to user 2 (106.3ms vs. 40.9ms) and from Figure~\ref{fig:correlation} we see user 1 also has a greater HRV range, shown by the ranges in the y-axes of the two graphs.  User 1 also has a grater range of timings for keystroke typing than user 2.

In terms of  bigram usage, not all of the top-10 bigrams  are used equally by any individual in any 5-minute typing window. We observe in Table~\ref{tab:timings} that user 1  has an obvious timing peculiarity when  typing the RE bigram, taking an average of 475ms across all recorded typing   compared to less than 100ms for almost all  other top-10 bigrams. When we removed the RE bigram from calculating keystroke dynamic deviations this improved the correlation but only marginally.  A similar observation was made when removing other  top-10 bigrams individually from the overall keystroke timing representation. Even for using just 1 of the bigrams, for both users, the correlation between HRV variance and keystroke timing variance against respective baselines, was weak.
What this tells us  is that we cannot take the observed  timing information compared against the baselines of only the top-10 most frequently occurring bigrams or even any group of them as the sole feature for characterising typing information and that we need a richer representation of timing information drawn from across more of the bigrams.

%


\section{Conclusions}

In this paper, we explored the relationship between  keystroke dynamics and heart-rate variability to find a relationship between a person's stress level as represented by changes in their HRV compared to their baseline, with variations in their keystroke timing information. We  used the Loggerman application to collect keystroke data for 2 participants and the Scosche RHYTHM24 band to collect HRV data for the same  participants for approximately the same amount of time. We presented the speed or latency at which each of the top-10 most frequently occurring bigrams from the English language were typed and how short-term HRV values measured over 5 minute windows which is the standard used for HRV, differed over a period of time for the users. 

We  observed from our analysis that there is little correlation between keystroke timing latency as represented by the top-10 bigrams, and HRV value changes yet we know from work elsewhere that keystroke timing information does correlate with user stress and cognitive load, as do variations in HRV. This suggests there is  scope for extracting alternative features from keystroke timing, other than timing of top-10 bigrams,  which might allow us to correlate with HRV data.  We are severely constrained here because whatever keystroke timing information is extracted from raw timing data, can only be from within a 5-minute window as that is the standard duration for calculating short-term HRV values~\cite{noauthor_chapter_nodate}.  We know that each individual has unique timing habits for their typing \cite{smeaton2021keystroke} and it may be that instead of choosing the same top-10 most frequently used bigrams from Norvig's analysis, a different subset might be appropriate for each user.

For future work, there is a spectrum of different platforms on which we could look into for collecting keystroke data. In this work we were limited to Apple Mac users due to limitations of the Loggerman application being available only on  OS/X.  So as to increase the participant base, we are capturing  keystroke data from mobile devices. HRV data could also be recorded for longer durations if even lesser intrusive wearable devices are used with longer battery life. 
Finally, of course we would like to get more participants to record for longer periods and possibly to have some of this in controlled environments. For example we could indirectly measure stress levels from keystroke dynamics for people working from home vs. working in an office environment, or we could measure mental state as students get closer to examination time.  

\vspace{0.5cm}
\noindent 
{\bf Acknowledgements:}
We are grateful to our participants for sharing their data with us.  This work was partly supported by Science Foundation Ireland (SFI) under Grant Number SFI/12/RC/2289\_P2, co-funded by the European Regional Development Fund

%

\bibliographystyle{splncs04} 
\bibliography{bibfile} 

\end{document}